\documentclass[preprint2]{aastex}

\usepackage{graphicx}
\usepackage{epstopdf}
\usepackage{color}
\usepackage{amsmath}

\slugcomment{published in Astrophys. J. 830 (2016) 145}

\shorttitle{Visible absorptions of C$_9$H$_9$ and C$_9$H$_5$}
\shortauthors{M. Steglich, S. Maity, \& J. P. Maier}

\begin{document}

\title{Visible absorptions of potential diffuse ISM hydrocarbons:\\ C$_9$H$_9$ and C$_9$H$_5$ radicals\\ \vspace{0.5cm} \small{published in \textit{Astroph. J.} 830 (2016) 145}}

\author{M. Steglich\altaffilmark{1}, S. Maity\altaffilmark{2}, J. P. Maier}
\affil{Department of Chemistry, University of Basel, Klingelbergstrasse 80, 4056 Basel, Switzerland}

\altaffiltext{1}{present address: Molecular Dynamics Group, Paul Scherrer Institute, 5232 Villigen PSI, Switzerland}
\altaffiltext{2}{present address: Department of Chemistry, University of Bern, Freistrasse 3, 3007 Bern, Switzerland}

\begin{abstract}
The laboratory detection of previously unobserved resonance-stabilized C$_9$H$_5$ and C$_9$H$_9$  radicals in the supersonic expansion of a hydrocarbon discharge source is reported. The radicals are tentatively assigned as acetylenic-substituted cyclopentadienyl C$_9$H$_5$ and vinyl-substituted benzyl C$_9$H$_9$ species. They are found to feature visible absorption bands that coincide with a few very weak diffuse interstellar bands toward HD183143 and HD204827. 
\end{abstract}

\keywords{molecular data --- methods: laboratory: molecular --- ISM: lines and bands --- ISM: molecules}

\section{Introduction}

In the past two decades, the electronic spectra of a number of carbon chains and their ions could be obtained in the gas phase at low temperatures \citep[][and references therein]{zack14}. The object of the studies was to be able to compare the laboratory data with astronomical measurements on the diffuse interstellar bands \citep[DIBs; e.g.,][]{herbig95}. The systems were relevant because radio astronomy has unambiguously identified polar carbon chains in dense clouds\footnote{www.astro.uni-koeln.de/cdms/molecules} and thus the absorptions of those systems which have their electronic transitions in the visible and near-infrared were sought. The laboratory studies were successful for a number of bare carbon chains and their cations and anions as well as those containing a few hydrogen atoms. It could be shown that none of their absorptions correspond to the known DIBs, which number up to five hundred by now \citep{snow06}. The general conclusion drawn was that such systems comprising up to a dozen or so carbon atoms cannot be responsible for the stronger DIBs, but larger systems like those with an odd number of at least fifteen carbon atoms remain viable candidates because they possess intense electronic transitions in the DIB range \citep{maier04}. Their spectra are known in the condensed phase \citep{forney96, wyss99} but not in gas yet. Only recently, the fullerene cation C$_{60}^+$ was confirmed to carry five DIBs between 9349 and 9633\,\AA\ \citep{campbell15, campbell16, walker15}. 

Analysis of the apparent rotational profiles of a few DIBs with resolved structure has lead \citet{huang15} to conclude that their most likely carriers are smaller molecules with five to eight heavy atoms. The initially evident systems, like C$_6$H and others, have already been studied, albeit with negative results \citep{motylewski00}. Therefore, we began to search for systems which have been omitted from our laboratory investigations until now. The first such example was the H$_2$C$_7$H$^+$ chain, the protonated form of the neutral cumulene, H$_2$C$_7$. The latter is a known constituent in dense clouds. Unfortunately, the absorption of H$_2$C$_7$H$^+$ falls in the region of a helium line precluding a conclusion \citep{rice15}.

However, it could be argued that cations of smaller molecules in diffuse interstellar space are subject to electron recombination and hence the neutral systems are of more relevance. In our studies of the electronic spectra of hydrocarbon radicals, also for interest as intermediates in combustion processes, we obtained spectra of a number of species of the type C$_n$H$_m$ \citep[e.g.,][]{maity15a, maity15b}, i.e. carbonaceous molecules containing more than only two or three hydrogens. Some of these fit into the category Oka's work has sought: a prolate structure and a small number of heavy atoms. In addition, hydrogenated carbonaceous matter is known to exist in the diffuse interstellar medium as can be concluded from the ratio of the 3.4 to 3.3 $\mu$m absorptions \citep[e.g.,][]{steglich13}.

In this article, we report the electronic spectra of previously unobserved radicals with molecular formula C$_9$H$_5$ and C$_9$H$_9$ that seem to match a few weak DIBs found by \citet{hobbs08,hobbs09}. The approach used to obtain the gas phase spectra was a multiphoton laser ionization technique applied on a molecular beam, which contains the target species under conditions as in diffuse interstellar clouds, i.e. with rotational temperatures of around 10\,K.

\section{Experiment}

Details of the experimental setup are discussed elsewhere \citep{maity15a}. The C$_9$H$_9$ and C$_9$H$_5$ radicals were produced along with other molecules in a pulsed electrical discharge of an organic precursor. The latter was a decomposition product of 0.1\% 1,6-heptadiyne (C$_7$H$_8$) in He. The decomposition occured during storage of the prepared gas bottle at 20$\degr$C for at least 24 hours; a discharge of the freshly prepared gas mixture did not produce the target species. The molecular beam was created by skimming the supersonic expansion (5 bar backing pressure) at about 30\,mm distance from the exit of the discharge source. Spectra in the 4200--7000\,\AA\ range were recorded using a resonant two-color two-photon ionization scheme. Photons with tunable energy, generated by an optical parametric oscillator (OPO; 20\,Hz, $\sim$5\,ns, 0.5--3 mJ/pulse, bandwidth $= 1$\,\AA, $\lambda$ uncertainty $= 5$\,\AA) or by a higher-resolution dye laser (20\,Hz, $\sim$5\,ns, 0.5--2 mJ/pulse, $\Delta\lambda = 0.02$\,\AA), counterpropagate the molecular beam and excite electronic states upon resonance. Ionization is achieved 0--5\,ns later by the 5th harmonic output (2128\,\AA, 1.5\,mJ/pulse) of the same Nd:YAG laser that pumps the OPO or dye laser, respectively. Resulting ions were extracted into a time-of-flight spectrometer and detected by a micro-channel plate.

The experiment is mass-selective, but a number of isomers is produced. The identification of the structures usually relies on comparison to quantum-chemical calculations of ionization energies, vibrational and electronic state energies, as well as vibrational progressions and rotational profiles. Those were obtained using density functional theory (DFT) and time-dependent DFT (TDDFT) as implemented in Gaussian09 \citep{frisch13}. The B3LYP functional \citep{becke88,lee88} was applied in conjunction with the cc-pVTZ basis set \citep{dunning89,woon93}. Vibrational progressions and rotational profiles were simulated with Gaussian09 and PGOPHER \citep{western10}.

\section{Results \& Discussion}

\begin{figure}\begin{center}
\epsscale{1.0} \plotone{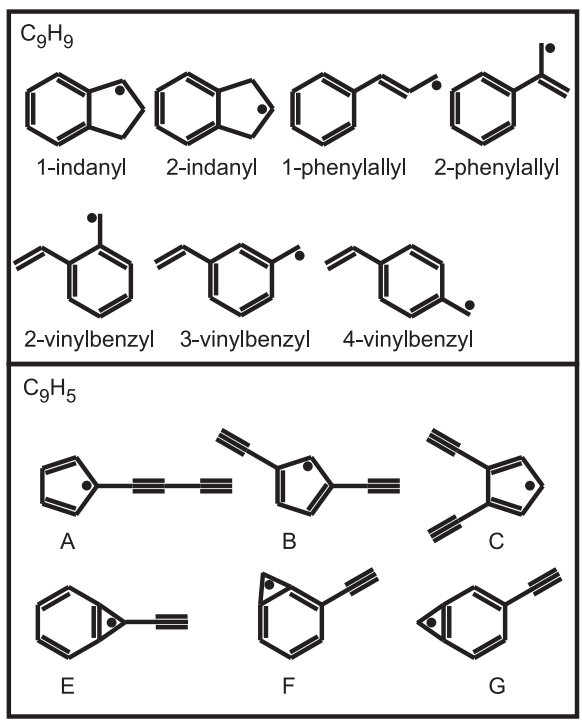} \caption{Structural formulae of the different radicals discussed in the text.} \label{fig_structures}
\end{center}\end{figure}

Resonance-stabilized hydrocarbon radicals (RSRs) are relatively long-lived species that can build up in high concentrations in energetic environments, such as plasmas, flames, planetary atmospheres, and interstellar space. Well known examples are the allyl C$_3$H$_5$, propargyl C$_3$H$_3$, and benzyl C$_7$H$_7$ radicals, which are key components in benzene and soot formation in hydrocarbon combustion. The higher stability of these molecules compared to other isomers is caused by a delocalization of the unpaired electron. This effect can be even more pronounced in larger entities. Recent laser spectroscopic investigations report on a number of such radicals in supersonic jets \citep{chalyavi11,chalyavi12,kidwell13,kidwell14,maity15a,maity15b,oconnor13,oconnor15,reilly08,reilly09a,reilly09b,sebree10a, sebree10b,sebree11,troy09,troy11,troy14,tsuge06}. Several of them could be recognized among the discharge products of the (partly decomposed) 1,6-heptadiyne precursor, e.g., cis- and trans-1-vinylpropargyl C$_5$H$_5$ \citep{reilly09b}, 1,4-pentadienyl C$_5$H$_7$ \citep{chalyavi11}, benzocyclopropenyl C$_7$H$_5$ \citep{maity15b}, benzyl C$_7$H$_7$ \citep{foster89}, 1-phenylpropargyl C$_9$H$_7$ \citep{reilly09a}, and 1-indanyl C$_9$H$_9$ \citep{troy09,maity15a}. We furthermore observed the triplet chains HC$_5$H and HC$_7$H \citep{steglich15,ding03} as well as previously unidentified species on masses corresponding to C$_7$H$_5$, C$_7$H$_7$, C$_7$H$_9$, C$_8$H$_9$, C$_9$H$_5$, and C$_9$H$_9$. The new C$_9$H$_5$ and C$_9$H$_9$ electronic absorptions are presented in the following, the strongest of which seem to agree with a few very weak DIBs. Structural formulae of C$_9$H$_5$ and C$_9$H$_9$ isomers are given in Fig. \ref{fig_structures}.

\subsection{C$_9$H$_9$}

\begin{figure*}\begin{center}
\epsscale{2.1} \plotone{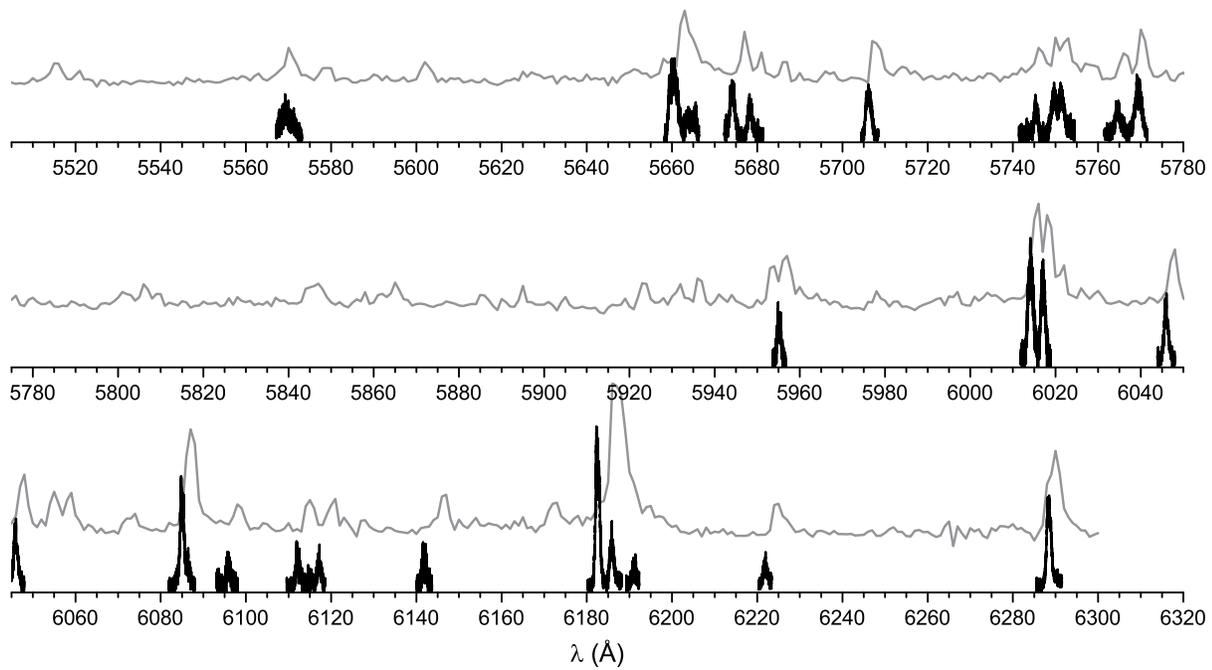} \caption{Absorption bands of C$_9$H$_9$ radicals. The OPO scan (1\,\AA\ bandwidth, 5\,\AA\ uncertainty) is displayed in gray and the higher-resolution scans (0.02\,\AA\ resolution) are the black traces.} \label{fig_C9H9all}
\end{center}\end{figure*}

\begin{figure}\begin{center}
\epsscale{1.0} \plotone{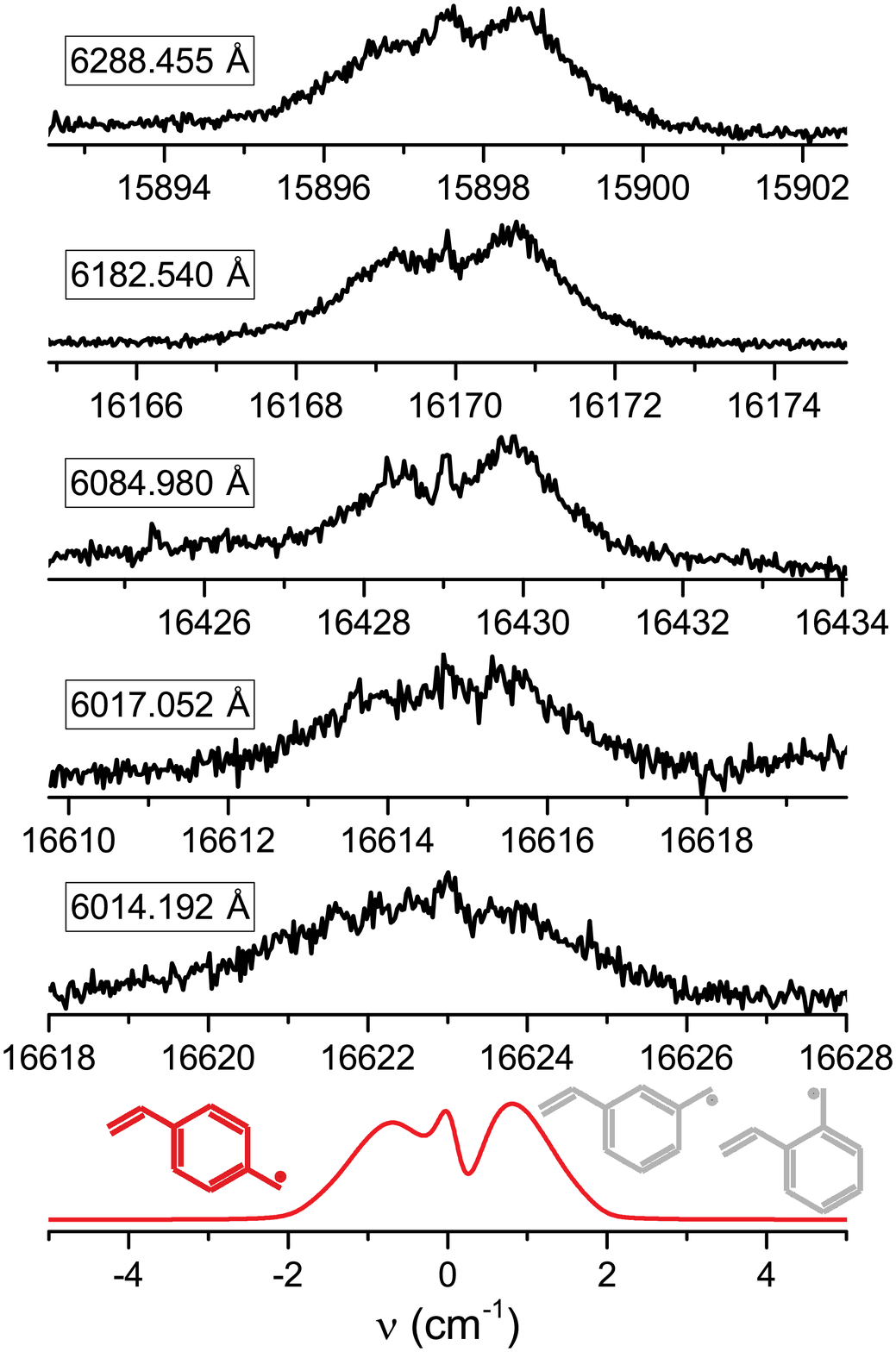} \caption{Higher-resolution observations of C$_9$H$_9$ absorptions. The bottom panel (red trace) is the calculated a-type rotational profile of 4-vinylbenzyl based on optimized geometries in D$_0$ and D$_1$ ($T_{rot} = 10$\,K). The meta and ortho isomers (3- and 2-vinylbenzyl; gray structures) have similar contours.} \label{fig_C9H9detail}
\end{center}\end{figure}

The most stable C$_9$H$_9$ isomer is the 1-indanyl radical. Its origin band at 4725\,\AA\ \citep{troy09,maity15a} is the strongest feature in the D$_1$$\leftarrow$D$_0$ progression. It was observed in this experiment, but about fifty times weaker than the most intense band of the new absorption system, which is displayed in Figs. \ref{fig_C9H9all} and \ref{fig_C9H9detail} and summarized in Table \ref{table_C9H9all}. The next stable isomer is the 1-phenylallyl radical, with a ground state energy calculated 0.17\,eV above 1-indanyl and a D$_1$$\leftarrow$D$_0$ origin measured at 5206\,\AA\ \citep{troy11, sebree11}. This one was not detected here, albeit the photon energies of the used lasers would have allowed it. The ground state of the second indanyl radical, i.e. 2-indanyl, is calculated 0.45\,eV higher than 1-indanyl. It should absorb more to the blue around 3000\,\AA, outside of the scan range. Located at comparable energies are the ground states of the 4-, 3-, and 2-vinylbenzyl radicals\footnote{The configurational isomerisms of 2- and 3-vinylbenzyl will be neglected in the following discussion, which will treat only the trans isomers. The cis forms are calculated to be less stable by 0.02--0.07\,eV. In the discharge experiments performed by \citet{troy11} and \citet{sebree11}, only the more stable trans isomer of 1-phenylallyl could be detected.} (0.44\,eV, 0.53, and 0.60\,eV) and of the 2-phenylallyl radical (0.53\,eV). The  D$_1$$\leftarrow$D$_0$ transitions of the vinylbenzyl radicals are predicted to cause adiabatic absorptions between 2.20 and 2.40\,eV, in proximity to the observed band system (1.97--2.23 eV). The 2-phenylallyl radical, on the other hand, should absorb somewhat further to the blue at 2.66\,eV. By scanning the energy of the ionizing laser (using another OPO between 2200 and 2400\,\AA), the ionization energy (IE) of the 6288\,\AA\ band carrier was determined to be $(7.3 \pm 0.1)$\,eV. The adiabatic IE of the 2-phenylallyl radical is calculated a bit too far to the blue (7.71\,eV) to justify a convincing assignment. The corresponding values of the vinylbenzyl radicals (6.81--7.07\,eV), especially the meta one (7.07\,eV), provide a slightly better match. The C$_9$H$_9$ isomers at next higher ground state energies (0.72--0.91\,eV above 1-indanyl) are aliphatic five- and six-membered carbon rings fused together, i.e., they have the same carbon skeleton like indanyl, but two hydrogens attached on one of the hexagon carbons, destroying the aromaticity. These structures have predicted D$_1$$\leftarrow$D$_0$ transitions close to the observed bands, too (2.04--2.40\,eV adiabatic), but their calculated ionization energies are considerably lower (6.08--6.65\,eV adiabatic). They can rather be excluded as potential carriers of the observed band system, leaving the three vinylbenzyl radicals and, maybe, the 2-phenylallyl radical as most likely explanation. This assignment receives additional support in view of the detection of other monocyclic aromatics in the same experiment, especially the benzyl one.

The measured rotational profiles feature a Q branch centered between the P and R ones (see Fig. \ref{fig_C9H9detail}). Assuming an asymmetric top rotor, this suggests a transition without a major geometry change that is of a-type, i.e., the dipole moment changes along the axis of smallest moment of inertia. This is in line with the suggested vinylbenzyl carriers, whose D$_1$$\leftarrow$D$_0$ transitions are predicted to be mainly a-type. A simulation of the rotational contour of the para isomer based on calculated geometries in D$_0$(A$''$) and D$_1$(A$''$) at $T_{rot} = 10$\,K is shown in the bottom panel of Fig. \ref{fig_C9H9detail}. Apart from a somewhat stronger broadening of the bands at higher energies, the agreement between simulated and observed profiles is reasonable in terms of widths and positions of the branches. The meta- and ortho-vinylbenzyl isomers have similar profiles predicted. The simulated rotational profile of the  D$_1$(B)$\leftarrow$D$_0$(A) transition of 2-phenylallyl (not shown here) provides an equally good match due to a mixed b and c character (about 1:1). A minor geometry change upon electronic transition should go along with a vibrational progression that is dominated by the origin band. Franck-Condon simulations indicate that this is rather true for the vinylbenzyl radicals than for 2-phenylallyl. Nevertheless, the calculated vibrational frequencies in the excited state and Franck-Condon intensities for such big molecules are usually too unreliable for a confident assignment of the observed vibrational progression (assuming that several carriers contribute). The calculated adiabatic D$_1$$\leftarrow$D$_0$ origin wavelengths and oscillator strengths are 5280\,\AA\ / $f=2\times10^{-3}$ (para-vinylbenzyl), 5160\,\AA\ / $f=4\times10^{-4}$ (meta), and 5640\,\AA\ / $f=1\times10^{-3}$ (ortho). Considering these values, the two strong bands at 6288 and 6183\,\AA\ may be assigned to the origin bands of ortho- and para-vinylbenzyl. Bands more to the blue could belong to vibronic excitations of all three vinylbenzyls and the 2-phenylallyl radical (4670\,\AA\ / $f=3\times10^{-3}$). These features remain underdetermined until further experiments are performed, e.g., hole burning or use of specific precursors.

Table \ref{table_C9H9detail} compares the five strongest bands of the recorded C$_9$H$_9$ spectrum with tabulated DIB data from HD204827 and HD183143 \citep{hobbs08,hobbs09}. The positions of the laboratory bands at 6182 and 6085\,\AA\ are within 0.2\,\AA\ of two very weak DIBs (equivalent width W$<$10\,m\AA). Band widths (FWHM) of around 1\,\AA\ are similar in both cases. The relative intensities do not correlate so well, but one has to keep in mind that the laboratory bands could stem from different isomers and that extracting equivalent widths from such weak interstellar features comprises some uncertainties. Furthermore, the density of DIBs, especially weak ones, in this wavelength region is high. Unresolved overlaps are a possibility. This certainly seems to be a problem in the case of a strong (W = 460...1880\,m\AA) DIB at 6284\,\AA, which is superposed by two to three weak features on its red shoulder that are nearby the 6288\,\AA\ laboratory band. Finally, very weak and narrow DIBs in HD204827 at 6019 and 6014\,\AA\ are comparable to the experimental bands at 6017 and 6014\,\AA. The same DIBs seem to appear in HD183143 around 6017 and 6015\,\AA, but were not tabulated\footnote{See full Fig. 11 in \citet{hobbs08,hobbs09}, available at http://dibdata.org.}. A more detailed comparison of band shapes is hardly feasible as the signal-to-noise ratio of the astronomical data is too low.

How do the known electronic absorptions of the more stable C$_9$H$_9$ isomers compare with Hobbs et al.'s DIB data? The strongest band in the D$_1$$\leftarrow$D$_0$ progression of 1-indanyl is the origin. It is at 4725\,\AA\ and has a width of 2\,\AA. A nearby DIB is at 4727\,\AA\ (W = 284\,m\AA , FWHM $= 2.7$\,\AA\ in HD204827; W = 156\,m\AA , FWHM $= 3.1$\,\AA\ in HD183143). Because of a 1\,\AA\ position mismatch, the 1-indanyl origin band was rejected as possible carrier for the 4727\,\AA\ DIB \citep{troy09}. Considering the previously discussed coincidences of absorptions of other C$_9$H$_9$ isomers with DIBs that are about thirty times weaker than the 4727\,\AA\ one and in view of comparable oscillator strengths of the D$_1$$\leftarrow$D$_0$ transitions ($1-2\times10^{-3}$), it could be argued that the 1-indanyl orgin might be hidden underneath the strong DIB, with which it was compared. The first transition of the trans-1-phenylallyl radical is also dominated by its origin, centered at 5205\,\AA. The calculated oscillator strength ($1\times10^{-3}$) is about half of that of the corresponding 1-indanyl transition. No DIBs are listed at 5205\,\AA, but a very weak interstellar feature possibly matching the 1-phenylallyl absorption seems to be present in the spectral data of HD204827.

\subsection{C$_9$H$_5$}
\begin{figure*}\begin{center}
\epsscale{2.1} \plotone{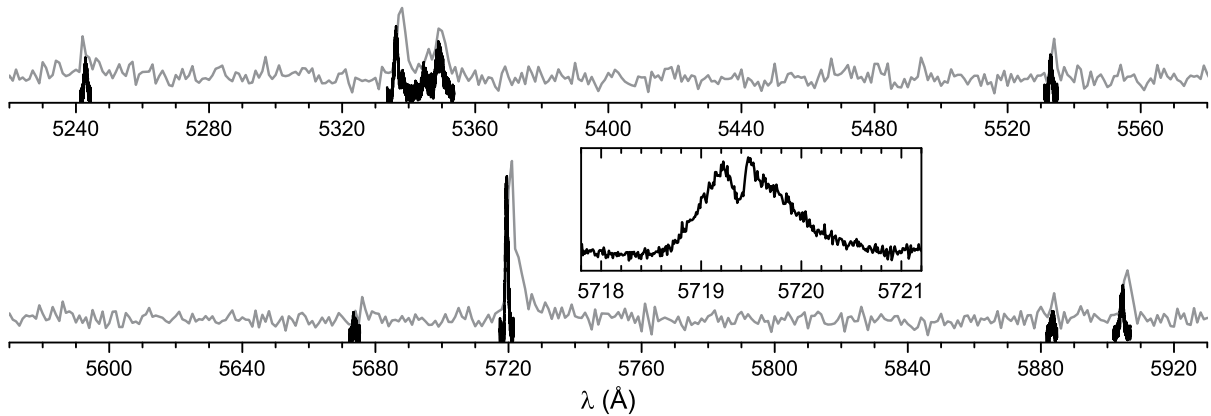} \caption{Absorption bands of C$_9$H$_5$ radicals. The OPO scan (1\,\AA\ bandwidth, 5\,\AA\ uncertainty) is displayed in gray and the higher-resolution scans (0.02\,\AA\ resolution) are the black traces. The inset is an expanded view of the strongest band.} \label{fig_C9H5all}
\end{center}\end{figure*}

\begin{figure}\begin{center}
\epsscale{1.0} \plotone{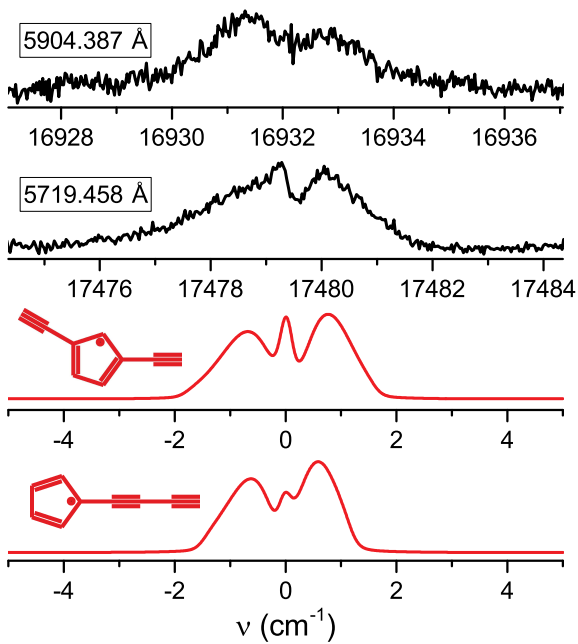} \caption{Higher-resolution observations of C$_9$H$_5$. The bottom panels (red traces) are D$_3 \leftarrow$D$_0$ rotational profile simulations of the lowest energy isomers ($T_{rot} = 10$\,K).} \label{fig_C9H5detail}
\end{center}\end{figure}

The absorption bands observed on $m/z = 113$, corresponding to C$_9$H$_5$, are plotted in Figs. \ref{fig_C9H5all} and \ref{fig_C9H5detail}. Band positions, widths and intensities are given in Table \ref{table_C9H5}. The strongest band at 5719.46\,\AA\ (FWHM = 0.99\,\AA) matches a weak DIB in HD183143 \citep[5719.63\,\AA; W = 21.7\,m\AA; FWHM = 0.92\,\AA;][]{hobbs09} and HD204827 \citep[5719.48\,\AA; W = 16.7\,m\AA; FWHM = 0.69\,\AA;][]{hobbs08}. The first band of the observed progression at 5904.39\,\AA\ (FWHM = 1.17\,\AA) is three times weaker than the strongest one. It also seems to coincide with a DIB, which is at 5904.63\,\AA (W = 4.6\,m\AA; FWHM = 0.85\,\AA) in HD204827. The same DIB appears in the spectrum of HD183143, but was not tabulated.

Several C$_9$H$_5$ isomers can be excluded as carriers of the experimental band system due to differences between computed transition or ionization energies and measured values. Among these are all linear carbon geometries, which are at least 1\,eV above the ground state structure. In addition, the simulated rotational profiles of the chains are difficult to bring into agreement with the observed contours. All cyclic isomers, i.e., those forming one big carbon ring, have no transitions predicted at the right energies or are too far away from the ground state energy of the most stable isomer. Two classes of molecules remain as plausible candidates. The lower energy class is made of a cyclopentadienyl ring (C$_5$H$_5$), which has two acetylenic --C$\equiv$C-- units attached, generating three different isomers. The one with lowest ground state energy is buta-1,3-diynyl-cyclopentadienyl H--C$\equiv$C--C$\equiv$C--(C$_5$H$_4$) (\textbf{A}). The other two are 1,3- and 1,2-diethynyl-cyclopentadienyl H--C$\equiv$C--(C$_5$H$_3$)--C$\equiv$C--H (\textbf{B} and \textbf{C}), which are slightly less stable by 0.16 and 0.25\,eV, respectively. The computed vertical excitation energies of allowed transitions in C$_9$H$_5$-\textbf{C} are about 0.7\,eV away from the observed absorptions, which were measured between 2.10 and 2.36\,eV. Isomers \textbf{A} and \textbf{B} have transitions calculated much closer at 2.57 and 2.71\,eV (adiabatic energies) with oscillator strengths $f=0.07$ and $f = 0.1$, respectively. The simulated rotational contours in their D$_3 \leftarrow$D$_0$ transitions are compared to the measured profiles of the 5904 and 5719\,\AA\ bands in Fig. \ref{fig_C9H5detail}, demonstrating a reasonable agreement. Weak Franck-Condon activities with strong origin bands predicted for both isomers are compatible with the absorption spectrum. The other class of possible carrier molecules is formed by attaching a single --C$\equiv$C-- unit to benzocyclopropenyl C$_7$H$_5$, generating H-C$\equiv$C-(C$_7$H$_4$). This gives also three different isomers (\textbf{E, F, G}), which are less stable than \textbf{A} by 0.62--0.85\,eV and have transitions predicted within 0.5\,eV (adiabatic energy; $f=4...16\times10^{-3}$) of the observed system. Their simulated rotational profiles (D$_1\leftarrow$D$_0$) are comparable to those of \textbf{A} and \textbf{B}, so they cannot be ruled out as possible band carriers with certainty. 

\section{Summary}
Visible absorption bands of previously unobserved C$_9$H$_5$ and C$_9$H$_9$ species were detected in the supersonic expansion of a hydrocarbon discharge source. The C$_9$H$_5$ radicals are tentatively assigned to resonance-stabilized buta-1,3-diynyl-cyclopentadienyl H--C$\equiv$C--C$\equiv$C--(C$_5$H$_4$) and 1,3-diethynyl-cyclopentadienyl H--C$\equiv$C--(C$_5$H$_3$)--C$\equiv$C--H. Isomers of vinyl-substituted benzyl probably carry the C$_9$H$_9$ absorptions. These assignments need to be verified in dedicated experiments, e.g., by applying carrier formation routes that use a more direct top-down approach (discharge of closely related precursors) or by probing the electronic ground state. 

The strongest C$_9$H$_9$ bands are found to coincide with very weak DIBs toward HD183143 and HD204827. Assuming that the laboratory bands are due to vinyl-substituted benzyl isomers the respective column densities are estimated between 1 and 3$\times 10^{13}$\,cm$^{-2}$, using $N = \dfrac{4 \epsilon_0 m_e c^2 W}{e^2 \lambda^2 f}$. Two further weak DIBs seem to match laboratory bands of C$_9$H$_5$, supposedly acetylenic substituted cyclopentadienyl. The corresponding column densities would be $0.5-1\times 10^{12}$\,cm$^{-2}$. Unequivocal confirmation of the presence of C$_9$H$_9$ and C$_9$H$_5$ in diffuse interstellar space requires astronomical observations with higher signal-to-noise ratio.


\acknowledgments

This work has been funded by the Swiss National Science Foundation (Project 200020-140316/1).

\begin{deluxetable}{ccc}
\tablecaption{Absorption bands of C$_9$H$_9$ radicals.}
\tablehead{
\colhead{$\lambda$ (\AA)\tablenotemark{a}}          &\colhead{Intens.\tablenotemark{b}}      &  \colhead{FWHM (\AA)} 
}
\startdata
6288.455	&	0.55	&		1.2				\\
6221.946	&	0.21	&		1.36			\\
6191.115	&	0.16	&		1.23			\\
6185.875	&	0.3		&		1.12			\\
6182.54	&	1		&		1.15			\\
(6173)		&	0.17	&		\nodata		\\
6141.794	&	0.25	&		1.13			\\
6117.138	&	0.22	&		1.01			\\
6112.176	&	0.21	&		1.17			\\
6095.904	&	0.18	&		1.19			\\
6086.4		&	0.18	&		1.14			\\
6084.98	&	0.66	&		1.09			\\
(6074)		&	0.13	&		\nodata		\\
(6059)		&	0.25	&		\nodata		\\
(6055)		&	0.25	&		\nodata		\\
6045.968	&	0.35	&		1.17			\\
(6022)		&	0.25	&		\nodata		\\
6017.052	&	0.58	&		1.23			\\
6014.192	&	0.66	&		1.42			\\
(5936)		&	0.19	&		\nodata		\\
(5932)		&	0.14	&		\nodata		\\
(5923)		&	0.17	&		\nodata		\\
(5895)		&	0.14	&		\nodata		\\
(5865)		&	0.16	&		\nodata		\\
(5847)		&	0.16	&		\nodata		\\
(5806)		&	0.15	&		\nodata		\\
5769.415	&	0.34	&		1.74			\\
5764.688	&	0.19	&		1.61			\\
(5757)		&	0.13	&		\nodata		\\
5751.415	&	0.29	&		1.36			\\
5749.611	&	0.29	&		1.39			\\
5745.425	&	0.23	&		0.95			\\
5706.138	&	0.27	&		1.48			\\
(5687)		&	0.13	&		\nodata		\\
5678.319	&	0.16	&		1.23			\\
5674.143	&	0.28	&		1.29			\\
(5668)		&	0.1		&		\nodata		\\
5660.386	&	0.45	&		2.55			\\
(5660)		&	0.12	&		\nodata		\\
(5658)		&	0.13	&		\nodata		\\
(5602)		&	0.13	&		\nodata		\\
5571.262	&	0.1		&		0.83			\\
5569.298	&	0.2		&		2.48			\\
(5515) 	&	0.13	&		\nodata		\\
\enddata
\label{table_C9H9all}
\tablenotetext{a}{Band center (higher-resolution scan with dye laser); accuracy of calibration is 0.02\,\AA. Values in brackets are uncalibrated OPO wavelengths ($\pm 5$\,\AA).}
\tablenotetext{b}{Relative peak intensity; absolute error of ca. 0.05.}
\end{deluxetable}

\begin{deluxetable}{ccccccccc}
\tablecaption{The strongest C$_9$H$_9$ absorptions compared to DIBs.}
\tablehead{
\multicolumn{3}{c|}{laboratory} & \multicolumn{3}{c|}{HD183143\tablenotemark{a}} & \multicolumn{3}{c}{HD204827\tablenotemark{b}} \\
\colhead{$\lambda$\tablenotemark{c}}          &	\colhead{Intens.\tablenotemark{d}}      &  \colhead{FWHM}  &	 \colhead{$\lambda$}	& \colhead{$W$}	&	\colhead{FWHM}  &	\colhead{$\lambda$}	& \colhead{$W$} & \colhead{FWHM} \\
\colhead{(\AA)}	& \colhead{}	& \colhead{(\AA)} & \colhead{(\AA)} &  \colhead{(m\AA)} & \colhead{(\AA)} 	& \colhead{(\AA)} & \colhead{(m\AA)} & \colhead{(\AA)}
}
\startdata
6288.455	&	0.55	&		1.2		& (6289.55)\tablenotemark{e}	& (19.2)	& (1.63)		& (6287.59)\tablenotemark{e}	& (13.9)	& (0.51)	\\
6182.54	&	1		&		1.15	& 6182.78							& 4.5		& 0.65			& 6182.58							& 6.4						& 1.04	\\
6084.98	&	0.66	&		1.09	& 6085.09							& 8.9		& 1.29			& 6084.94							& 6.8						& 0.82	\\
6017.052	&	0.58	&		1.23	& ?\tablenotemark{f}			&\nodata	&\nodata		& 6019.32							& 4.2						& 0.79	\\
6014.192	&	0.66	&		1.42	& ?\tablenotemark{f}			&\nodata	&\nodata		& 6014.81							& 3.4						& 0.73	\\
\enddata
\label{table_C9H9detail}
\tablenotetext{a}{\citet{hobbs09}.}
\tablenotetext{b}{\citet{hobbs08}.}
\tablenotetext{c}{Band center; accuracy of calibration is 0.02 \AA.}
\tablenotetext{d}{Relative peak intensity; absolute error of ca. 0.05.}
\tablenotetext{e}{Overlap with a strong DIB.}
\tablenotetext{f}{Nearby weak DIBs appear in the spectrum, but are not tabulated.}
\end{deluxetable}

\begin{deluxetable}{ccc}
\tablecaption{Absorption bands of C$_9$H$_5$ radical(s).}
\tablehead{
\colhead{$\lambda$\tablenotemark{a} (\AA)}	&	\colhead{Intens.\tablenotemark{b}}	&	\colhead{FWHM (\AA)}	}
\startdata
5904.387	&	0.31	&	1.17		\\
5883.507	&	0.16	&	1.07		\\
5719.458\tablenotemark{c}	&	1		&	0.99		\\
5673.589	&	0.14	&	1.02		\\
5532.972	&	0.26	&	1.01		\\
(5494) 	&	0.14	&	\nodata 	\\
5349.215	&	0.32	&	2.49		\\
5344.861	&	0.19	&	1.73		\\
5338.116	&	0.15	&	1.47		\\
5336.225	&	0.45	&	1.28		\\
(5297) 	&	0.14	&	\nodata 	\\
5242.934	&	0.27	&	0.96		\\
\enddata
\label{table_C9H5}
\tablenotetext{a}{Band center (higher-resolution scan with dye laser); accuracy of calibration is 0.02\,\AA. Values in brackets are uncalibrated OPO wavelengths ($\pm 5$\,\AA).}
\tablenotetext{b}{Relative peak intensity; absolute error of ca. 0.05.}
\tablenotetext{c}{The strongest band agrees with a weak DIB in HD183143 (5719.63 \AA; W = 21.7 m\AA; FWHM = 0.92 \AA) \& HD204827 (5719.48 \AA; W = 16.7 m\AA; FWHM = 0.69 \AA).}
\end{deluxetable}

\end{document}